# Handedness-filter and Doppler shift of spin waves in ferrimagnetic domain walls

T. T. Liu[1], Y. Liu[1], Z. Jin[1], Z. P. Hou[1], D. Y. Chen[1], Z. Fan[1], M. Zeng[1], X. B. Lu[1], X. S. Gao[1], M. H. Qin[1,*], and J. –M. Liu[1,2]

[1]*Guangdong Provincial Key Laboratory of Quantum Engineering and Quantum Materials and Institute for Advanced Materials, South China Academy of Advanced Optoelectronics, South China Normal University, Guangzhou 510006, China*

[2]*Laboratory of Solid State Microstructures, Nanjing University, Nanjing 210093, China*

**[Abstract]** Excitation and propagation of spin waves inside magnetic domain walls has received attention because of their potentials in spintronic and communication applications. Besides wave amplitude and frequency, spin-wave has its third character: handedness, whose manipulation is certainly of interest. We propose in this Letter that the handedness of low energy spin-wave excitations can be controlled by tuning the net angular momentum $\delta_s$ in a ferrimagnetic (FiM) domain wall, attributing to the inequivalent magnetic sublattices. The results indicate that the spin-wave dispersion depends on both $\delta_s$ and wave handedness. For a positive (negative) $\delta_s$, a gapless dispersion is observed for the left-handed (righ-handed) spin waves, while a frequency gap appears for the right-handed (left-handed) spin waves. Thus a FiM wall could serve as a multifold filter of low energy spin-wave in which only spin waves with particular handedness can propagate. Furthermore, the energy consumption loss for spin-wave excitation in the wall is much lower than that inside the domain, while the group velocity is much faster too, demonstrating the advantages of domain walls serving as spin waveguides. Moreover, the current-induced spin-wave Doppler shift in the FiM wall is also revealed, and can be controlled by $\delta_s$. This work unveils for the first time the interesting spin-wave dynamics in FiM domain walls, benefiting future spin-wave applications.

---

Email: qinmh@scnu.edu.cn

Conventional semiconductor devices transmit and process information with electric current. However, charge flow causes power consumption due to the Joule heating, a highly concerned drawback for advanced devices [1,2]. Recently, spin-wave devices have received much attention as a plausible complement to conventional semiconductor electronics [3–6]. Instead of charge transport, spin wave is used as information carrier in these devices, which is basically free of the Joule heating [7–10]. Moreover, spin-wave devices allow an integration of memory and computing into a single hardware unit through combining nonvolatile storage capabilities of magnetic materials with high-frequency data processing in THz regime [11–13]. Thus, it could break the bottleneck of modern CPU and memory architectures and play an essential role in novel information processing [14].

As a matter of fact, spin-wave devices have been proposed, including the nanoscale neural network [15], transistors [6], logic gates [16] and multiplexers [7], where the information is encoded in amplitude, frequency, or phase respectively [14,17]. Considering the information processing, waveguide structures can guide spin-wave propagation without disturbances between neighboring channels [17–20]. For example, the easy-axis surface anisotropy has been proven as a waveguide from the surface spin-wave mode [17,21], while high requirement for the fabrication is needed technologically. Along this line, magnetic structures such as domain wall can be an alternative choice of waveguides [22,23], noting that these structures generally have small sizes and can be easily modulated through various methods.

Actually, magnetic domain walls as natural waveguides have been reported [24–29]. In ferromagnets, low frequency spin waves, non-excited due to the gapped bulk spin wave modes, can be generated inside domain walls and guided in curved geometries [24]. However, due to the time-reversal symmetry breaking, spin wave can be only right-circularly polarized in the standard polar coordinates for a spin system [14,30]. In addition, strong stray field in ferromagnets [31–37] can be disadvantageous for high-density device integration. In antiferromagnets, both left- and right-handed spin waves can be excited [38], besides their zero stray field and fast magnetic dynamics [38,39]. Interestingly, the spin wave dispersion inside an antiferromagnetic domain wall is also gapless where the group velocity of spin wave is much higher than that in ferromagnets [25]. Unfortunately, despite these advantages for spintronic applications, it is still challenging to experimentally control antiferromagnetic dynamics due to their zero net magnetization [40,41].

As a compromise, ferrimagnet would be a preferred candidate in the sense of trade-off if spin wave can be properly excited and propagated, and it has limited net magnetic moment but rapid dynamics in the vicinity of angular momentum compensation temperature ($T_A$) [42–44]. The magnetic states at $T_A$ can be effectively detected and addressed by conventional methods. Indeed, spin waves have a full polarization degrees of freedom in ferrimagnets where two inequivalent magnetic sublattices are coupled antiferromagnetically and thus can be effectively excited and controlled [37,42]. In fact, spin wave propagation in a one-dimensional system with the ferrimagnetic (FiM) uniform state (i.e. a FiM domain) has been addressed recently, approving the excitation of spin waves once the excitation frequency is sufficiently high [38]. It is on the contrary suggested that the spin wave excitation in a FiM system could be an interesting issue and potentially emergent phenomena can be expected.

More interestingly, a FiM system may also accommodate frequency gap for spin wave excitation and propagation [38], and this gap can be reduced if the spin wave is confined inside the domain wall as a waveguide. It comes to our attention that a spin wave always carries its sign of spin polarization, which is intrinsically related with the spin-wave handedness. In the other words, this implies that the existence of the frequency gap probably depends on the spin-wave handedness, noting a recent experiment on the switch of spin-wave handedness near $T_A$ in ferrimagnet [45,46]. Furthermore, it is known that the spin-wave dynamics in FiM domain can be modulated by current-induced spin-transfer torques (STT) [37], and similar property could be expected for spin wave channeled in a FiM wall. In short, it is said that the propagation of spin waves in a FiM wall may be modulated by not only the frequency but also the wave-handedness, a double-advantage that has never been reported in literature here and definitelydeserves for investigation.

In this work, we intend to address this issue and investigate theoretically the excitation and propagation of spin waves confined inside a FiM wall. Without losing the generality, we consider a two-dimensional ultrathin FiM film in the *xy* plane, whose magnetic structure is shown in Fig. 1. A FiM wall aligned along the *x*-axis separates the two uniform domains, and a sinusoidal excitation field is applied to excite the spin waves which propagate along the *x* direction. In this geometry, one needs to address the spin-wave dynamics inside the domain

(DO) and domain wall (DW) respectively.

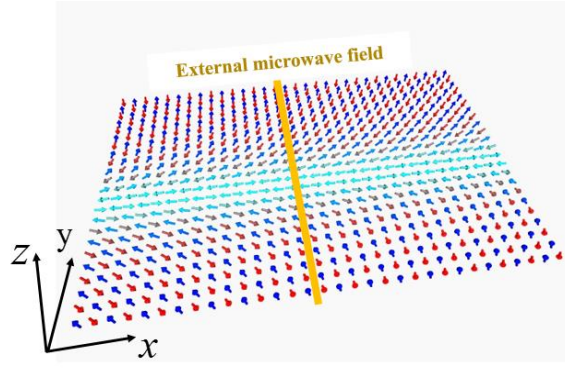

Fig. 1 (color online). Illustration of a FiM wall of the Bloch type. External microwave field is applied along the *x*-axis to excite the spin wave that propagates along the *x* direction too.

We introduce two unit vectors $\mathbf{m}_1$ and $\mathbf{m}_2$ to denote spin moments at the two sublattices, and define the staggered vector $\mathbf{n} = (\mathbf{m}_1 - \mathbf{m}_2)/2$ and averaged magnetization $\mathbf{m} = (\mathbf{m}_1 + \mathbf{m}_2)/2$. It is seen that the lattice would be ferromagnetic if $\mathbf{m}_1 = \mathbf{m}_2$ and antiferromagnetic if $\mathbf{m}_1 = -\mathbf{m}_2$. In addition, for the two sublattices, their gyromagnetic ratio and the Gilbert damping constant are respectively denoted as $\gamma_{1,2}$ and $\alpha_{1,2}$. The spin density of sublattice $i$ ($i = 1, 2$) is given by $s_i = M_i/\gamma_i$ with $\gamma_i = g_i \mu_B/\hbar$, where $M_i$ is the saturation magnetization, $g_i$ is the Landé factor, and $\mu_B$ is the Bohr magneton.

Following the standard procedure in earlier work [42,43], the Lagrangian density for a FiM lattice is given by

$$\mathcal{L} = -s\dot{\mathbf{n}} \cdot (\mathbf{n} \times \mathbf{m}) - \delta_s \mathbf{a}(\mathbf{n}) \cdot \dot{\mathbf{n}} - \mathcal{U}, \tag{1}$$

where $s = (s_1 + s_2)/2$ is the staggered spin density, $\delta_s = s_1 - s_2$ is the net angular momentum, and $\mathbf{a}(\mathbf{n})$ is the vector potential of a magnetic monopole satisfying $\nabla_\mathbf{n} \times \mathbf{a} = \mathbf{n}$. The potential energy density $\mathcal{U}$ is given by [40]

$$\mathcal{U} = \frac{A}{2}(\nabla \mathbf{n})^2 + \frac{a}{2}\mathbf{m}^2 + L\mathbf{m} \cdot \partial_x \mathbf{n} - \frac{K}{2}(\hat{z} \cdot \mathbf{n})^2, \tag{2}$$

where $A$ is the inhomogeneous exchange, $a$ is the homogeneous exchange, $L$ is the parity-breaking exchange term, and $K$ is the effective easy-axis anisotropy. The Rayleigh dissipation function is given by $R = s_\alpha \dot{\mathbf{n}}^2/2$ where $s_\alpha = \alpha_1 s_1 + \alpha_2 s_2$ is a phenomenological parameter quantifying the energy and spin loss due to the magnetic dynamics. For simplicity, we assume

$α_1 = α_2 = α$ and neglect the nonlocal dipolar interaction, considering the small net magnetization in the investigated systems.

In the presence of spin current, the dynamic governing equations obtained by solving the above equations for **n** and **m** are given as [38]

$$\dot{\mathbf{n}} = -\frac{1}{s}\mathbf{f_m} \times \mathbf{n} + \mathbf{T}_{STT}^{\mathbf{n}}, \tag{3a}$$

$$\dot{\mathbf{m}} = -\frac{1}{s}\mathbf{f}_n \times \mathbf{n} + 2\alpha\dot{\mathbf{n}} \times \mathbf{n} - \frac{\delta_s}{s}\dot{\mathbf{n}} + \mathbf{T}_{STT}^{\mathbf{m}}, \tag{3b}$$

where $\mathbf{f_n}$ and $\mathbf{f_m}$ are the effective fields associated with **n** and **m**, respectively. The effective fields with the aid of suitable Lagrangian multipliers subject to the aforementioned constraints leads to $\mathbf{f_m} = -\partial \mathcal{U}/\partial \mathbf{m} = -a\mathbf{m}$ and $\mathbf{f_n} = -\partial \mathcal{U}/\partial \mathbf{n} = A\mathbf{n} \times (\nabla^2\mathbf{n} \times \mathbf{n}) + K(\mathbf{n} \cdot \hat{z})\mathbf{n} \times (\hat{z} \times \mathbf{n})$. The STT modulating the **n** and **m** dynamics read $\mathbf{T}_{STT}^{\mathbf{n}} = -(b_j^+ \partial\mathbf{n}/\partial x + \beta b_j^- \mathbf{n} \times \partial\mathbf{n}/\partial x)/2$ and $\mathbf{T}_{STT}^{\mathbf{m}} = -b_j^- \partial\mathbf{n}/\partial x - \beta b_j^+ \mathbf{n} \times \partial\mathbf{n}/\partial x$, respectively, where $b_j^{\pm} = -\mu_B(P_1 g_1/M_1 \pm P_2 g_2/M_2)J_e/2e$ is the magnitude of adiabatic spin torque, $P_i$ is the spin polarization, $J_e$ is the current density, and $\beta$ is the nonadiabaticity.

Subsequently, we address the propagation of spin wave in a FiM wall. We ignore the damping term and consider a small fluctuation of the staggered vector **n** around the initial $\mathbf{n_0}$, $\mathbf{n} = \mathbf{n_0} + \delta\mathbf{n}(\mathbf{x}, t)$ with spin position **x** and time $t$. Then, the equations of motion for **n** and **m** are obtained as

$$\delta\dot{\mathbf{n}} = \frac{a}{s}\mathbf{m} \times \mathbf{n_0}, \tag{4a}$$

$$\dot{\mathbf{m}} = -\frac{\delta_s}{s}\dot{\mathbf{n}} - b_j^- \frac{\partial\mathbf{n}}{\partial x} - \beta b_j^+ \mathbf{n} \times \frac{\partial\mathbf{n}}{\partial x} - \frac{1}{s}\Big[A\big(\nabla^2\mathbf{n_0} \times \mathbf{n_0} + \nabla^2\mathbf{n_0} \times \delta\mathbf{n} + \delta\mathbf{n} \times \mathbf{n_0}\big) \\ + K_z\big((\mathbf{n_0} \cdot \hat{z})\hat{z} \times \mathbf{n_0} + (\mathbf{n_0} \cdot \hat{z})\hat{z} \times \delta\mathbf{n} + (\delta\mathbf{n} \cdot \hat{z})\hat{z} \times \mathbf{n_0}\big)\Big], \tag{4b}$$

We consider the Walker ansatz for a Bloch-type domain wall profile centered at $y = 0$ as shown in Fig. 1: $\mathbf{n_0} = [\text{sech}(y/\Delta), 0, \tanh(y/\Delta)]$ [25] with the domain wall width $\Delta = (A/K)^{1/2}$. Similarly, we transform the coordinates with mutually orthogonal basis vectors $\mathbf{j} = (0, 1, 0)$ and $\mathbf{i} = \mathbf{j} \times \mathbf{n_0} = [\tanh(y/\Delta), 0, -\text{sech}(y/\Delta)]$ to express the fluctuating fields $\delta\mathbf{n}$ and **m** in the plane perpendicular to $\mathbf{n_0}$. Considering that the monochromatic waves propagate along the $x$ direction, one obtains $\delta\mathbf{n} = [n_i(y)\mathbf{i} + n_j(y)\mathbf{j}]e^{i\mathbf{k}\cdot\mathbf{x} - i\omega t}$, $\mathbf{m} = [m_i(y)\mathbf{i} + m_j(y)\mathbf{j}]e^{i\mathbf{k}\cdot\mathbf{x} - i\omega t}$. By arranging Eq. (4) on the basis of **i** and **j**, one obtains

$$-i\omega s n_i(n_j) = a m_j(-m_i), \tag{5a}$$

$$-i\omega s m_i(m_j) = \delta_s i\omega n_i(n_j) - sb_j^- n_i(n_j)ik + s\beta b_j^+ ikn_j(-n_i)$$
$$-\left[A\frac{d^2 n_j(-n_i)}{dy^2} - Ak^2 n_j(-n_i) + K\left(\text{sech}^2(\frac{y}{\Delta}) - \tanh^2(\frac{y}{\Delta})\right)n_j(-n_i)\right], \quad (5b)$$

Eq. (5a) describes the dynamics of small variations $\delta\mathbf{n}$ in the **i** and **j** directions, and Eq. (5b) describes to the dynamics of the canted magnetization **m** in the **i** and **j** directions. Moreover, we eliminate $m_i$ and $m_j$ by coupling Eq. (5a) to Eq. (5b), and obtain

$$-\delta_s i\omega n_i + sb_j^- n_i ik - s\beta b_j^+ ikn_j$$
$$= -A\frac{d^2 n_j}{dy^2} + Ak^2 n_j - K\left(2\text{sech}^2(\frac{y}{\Delta}) - 1\right)n_j - \rho\omega^2 n_j, \quad (6a)$$

$$-\delta_s i\omega n_j + sb_j^- n_j ik + s\beta b_j^+ ikn_i$$
$$= A\frac{d^2 n_i}{dy^2} - Ak^2 n_i + K\left(2\text{sech}^2(\frac{y}{\Delta}) - 1\right)n_i + \rho\omega^2 n_i, \quad (6b)$$

where $\rho = s^2/a$ is the inertia of the dynamics, $d$ is lattice spacing. By defining a complex field as $\psi_\pm = n_i \mp in_j$ for right- (+) and left- handed (−) spin waves, Eq. (6a) and Eq. (6b) can be linearized and updated to

$$p^2\psi_\pm(\xi) = \left(-\frac{d^2}{d\xi^2} - 2\text{sech}^2\xi\right)\psi_\pm(\xi), \quad (7)$$

with $\xi = y/\Delta$ and $p^2 = (\rho\omega_\pm^2 \mp \delta_s\omega_\pm - Ak^2 \pm sb_j^- k)/K - 1$. Equation (7) has two types of solutions: the exact solutions for the domain state (marked by superscript "DO") and the wall state (marked by superscript "DW"), which correspond to $p = 0$ and $p = -i$, respectively [27,30]. Then, we obtain the dispersion relation given by

$$\omega_\pm^{DO} = \frac{\pm\delta_s + \sqrt{\delta_s^2 + 4\rho(Ak^2 \mp sb_j^- k + K)}}{2\rho}, \quad (8a)$$
$$\omega_\pm^{DO}(k=0) > 0, \text{ gapped}$$

$$\omega_\pm^{DW} = \frac{\pm\delta_s + \sqrt{\delta_s^2 + 4\rho(Ak^2 \mp sb_j^- k)}}{2\rho}$$
$$\omega_\pm^{DW}(k=0, \delta_s > 0) = \begin{cases} \delta_s/\rho, & \text{right-handed, gapped} \\ 0, & \text{left-handed, gapless} \end{cases}, \quad (8b)$$
$$\omega_\pm^{DW}(k=0, \delta_s < 0) = \begin{cases} 0, & \text{right-handed, gapless} \\ -\delta_s/\rho, & \text{left-handed, gapled} \end{cases}$$

where the so-called frequency gap is defined by $\omega_\pm(k=0)$.

Eqs. (8a) and (8b) describe the dispersion relation for spin wave propagation inside the domain and wall, respectively, where the + (−) sign corresponds to the right-circularly (left-circularly) polarized spin waves respectively.

It is noted that spin waves can be excited only when the frequency of the excitation field is larger than the frequency gap, if any. Obviously, there is a gap caused by the magnetic anisotropy for the spin wave inside the domain ($\omega^{DO} > 0$), and the gap magnitude also depends on the $\delta_s$ value and the handedness of the spin wave. For example, for a positive $\delta_s$, the frequency gap of the right-handed spin wave is larger than that of the left-handed spin wave.

Importantly, no frequency gap exists in the wall state for spin waves with the left handedness (gapless), as demonstrated in Eq. (8b). Thus, the left-handed spin waves can always propagate inside the wall. More interestingly, a nonzero gap still exists inside the wall for the right-handed spin waves, whose magnitude depends on the $\delta_s$ value, suggesting that low energy spin waves with the right-handedness cannot be excited. Thus, a FiM wall may serve as a waveguide for spin waves with a specific handedness, and as a filter for spin waves with the other handedness. Moreover, the frequency spectrum of the waveguide and/or filter can be easily controlled through various methods such as a magnetic field. These particular merits allow a wide application prospect of the FiM wall in future spin-wave devices.

Furthermore, in the absence of electric current, i.e. $b_j^- = 0$, the group velocity of spin wave is given by

$$v^{DO} = \frac{2Ak}{\sqrt{\delta_s^2 + 4\rho(Ak^2 + K)}}, \tag{9a}$$

$$v^{DW} = \frac{2Ak}{\sqrt{\delta_s^2 + 4\rho(Ak^2)}}, \tag{9b}$$

It is clearly shown that for any $k$, spin waves guided inside the wall propagate faster than that inside the domain, which demonstrates again the advantage of wall as the waveguide. Moreover, for $\delta_s = 0$ at $T_A$, $v^{DW}$ is independent of the frequency and vector, the same as the cases of antiferromagnets.

To verify the above analysis, we also perform numerical simulations based on the atomistic Landau-Lifshitz-Gilbert (LLG) equation. Here, the corresponding Hamiltonian is given by

$$\mathcal{H} = A_{sim}\sum_{i,j}\mathbf{S}_i \cdot \mathbf{S}_j - K_{sim}\sum_i (\mathbf{S}_i \cdot \hat{\mathbf{z}})^2, \tag{10}$$

where $A_{sim}$ is the exchange coupling constant between the nearest neighbors, $\mathbf{S}_i$ is the normalized magnetic moment at site $i$, and $K_{sim}$ is the easy-axis anisotropy along the $z$-axis.

In the presence of spin current in the current-in-plane geometry, the dynamics is investigated by solving the LLG equation

$$\frac{\partial \mathbf{S}_i}{\partial t} = -\gamma_i \mu_0 \mathbf{S}_i \times \mathbf{H}_{\text{eff},i} + \alpha_i \mathbf{S}_i \times \frac{\partial \mathbf{S}_i}{\partial t} - b_{J,i} \frac{\mathbf{S}_{i+1} - \mathbf{S}_{i-1}}{2d} - \beta b_{J,i} \mathbf{S}_i \times \frac{\mathbf{S}_{i+1} - \mathbf{S}_{i-1}}{2d}, \quad (11)$$

where $\mathbf{H}_{\text{eff},i} = -(1/\mu_i)\partial \mathcal{H}/\partial \mathbf{S}_i$ is the effective field with the magnetic moment $\mu_i$, and $b_{J,i} = -(g_i P_i \mu_B/2eM_i)J_e$ is the magnitude of adiabatic STT.

The system size is 1000 nm × 200 nm × 7 nm with a cell size of 2 nm × 2 nm × 7 nm. The absorbing boundary conditions are considered in the simulations. The initial magnetization state was obtained by relaxation with a Bloch-type domain-wall configuration for sufficiently long time. The external field $\mu_0\mathbf{H} = \mu_0 H_0(0, \pm\sin 2\pi ft, -\cos 2\pi ft)$ is applied to excite spin waves. By taking the well-known rare-earth transition-metal (RE-TM) ferrimagnet GdFeCo as an example, we use the following simulation parameters [47]: $A_{sim} = 15 \times 10^{-12}$ J/m, $K_{sim} = 2 \times 10^5$ J/m$^3$, $\mu_0 H_0 = 10$ mT, the current density $J_e = 1 \times 10^{13}$ A/m$^2$, the spin polarization $P_{RE} = 0.1$, $P_{TM} = 0.4$, and the gyromagnetic ratio $\gamma_{RE} = 1.76 \times 10^{11}$ rad/sT and $\gamma_{TM} = 1.936 \times 10^{11}$ rad/sT (correspond to the Landé factors $g_{RE} = 2$ and $g_{TM} = 2.2$, respectively). We consider the damping constant $\alpha_{RE} = \alpha_{TM} = 0.003$, and the nonadiabaticity $\beta = 0.03$. The magnetic moments $M_{RE}$ and $M_{TM}$ are listed in Table I.

TABLE I. Parameters chosen for the simulation.

| Parameter | 1 | 2 | 3 |
|---|---|---|---|
| $M_{TM}$ (kA/m) | 950.8 | 1100 | 1280 |
| $M_{RE}$ (kA/m) | 938 | 1000 | 1090 |
| $\delta_s$ (10$^{-6}$ Js/m$^3$) | −2.6 | 0 | 2.6 |

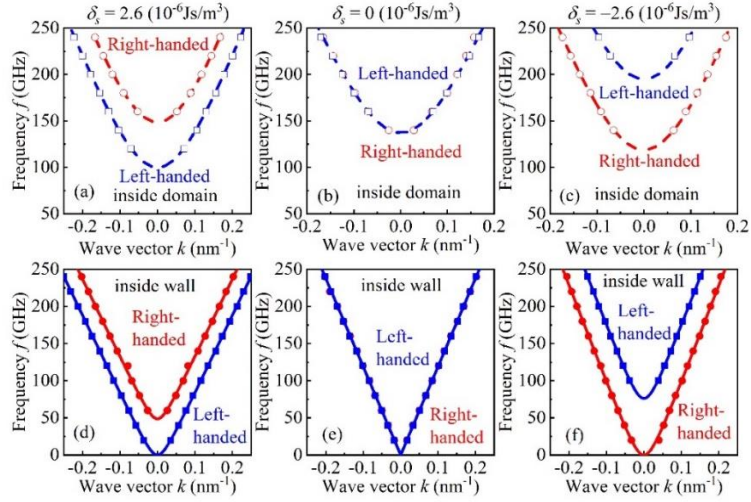

Fig. 2 (color online). Simulated (symbols) and calculated (lines) dispersion relations for spin waves inside domains at (a) $\delta_s > 0$, (b) $\delta_s = 0$ and (c) $\delta_s < 0$, and for spin waves inside walls at (d) $\delta_s > 0$, (e) $\delta_s = 0$ and (f) $\delta_s < 0$. There exists a nonzero gap if the frequency minimal is nonzero and the spin wave propagation is gapless if the frequency at $k = 0$ is zero.

Hereafter, we investigate the spin-wave dynamics in the absence of STT and address the effect of $\delta_s$ and handedness of spin wave. Fig. 2 presents the numerically simulated (symbols) and Eq. (8)-based calculated (lines) dispersion relations of the spin waves for various $\delta_s$ inside domain (a-c) and wall (d-f). The simulated data fit the calculations perfectly. For $\delta_s = 0$ at $T_A$, the dispersion relations for the right-handed and left-handed spin waves inside domain coincide with each other (Fig. 2(b)), similar to the case of antiferromagnets. Moreover, a frequency gap is observed and spin waves can be only excited if the frequency is higher than the gap (~137 GHz here). Interestingly, for a finite $\delta_s$, the gap significantly depends on the wave-handedness. Specifically, the gap of the right- (left-) handed spin wave is enlarged (diminished) by a positive $\delta_s$, as clearly shown in Fig. 2(a). In contrast, a negative $\delta_s$ suppresses the frequency gap of the right-handed spin wave and enhances that of the left-handed spin wave (Fig. 2(c)). Thus, the net angular momentum could be used to tune the gap for spin wave propagation inside domain, noting that the gap should be reduced and even eradicated to enhance the spin-wave velocity and reduce the exciting power.

For the wave inside the wall, at $\delta_s = 0$, the gapless dispersion is observed both for the right-handed and left-handed spin waves (Fig. 2(e)), the same as that of its antiferromagnetic counterpart. For a nonzero $\delta_s$, the spin-wave dispersion depends on the handedness, attributed to the inequivalent moments at two sublattices. The dispersion for the left-handed spin waves is gapless if $\delta_s > 0$, while a gap appears for the right-handed spin waves, as shown in Fig. 2(d). This behavior is rather different from the cases of ferromagnetic and antiferromagnetic domain walls where no frequency gap appears, regardless of the wave-handedness. It is thus suggested that a FiM wall could serve as a multifold filter for the low energy spin-wave in which only waves with particular handedness can be excited and propagated. Fig. 2(f) presents the spin-wave dispersions inside the wall for $\delta_s < 0$, which is gapless for the right-handed spin waves and gapped for the left-handed spin waves.

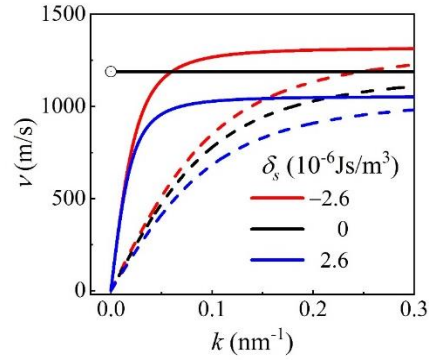

Fig. 3 (color online). The group velocity of spin waves inside wall (solid lines) and domain (dashed lines) of a FiM system for various $\delta_s$.

Subsequently, we tend to estimate the spin-wave velocity and excitation energy. From Eq. (9), the group velocity of spin wave depending on $k$ and $\delta_s$ is predicted, independent of the handedness. Fig. 3 presents the calculated $v^{DO}$ (inside domain, dashed lines) and $v^{DW}$ (inside wall, solid lines) as a function of $k$ for various $\delta_s$. It is seen that $v^{DW}$ is independent of $k$ for $\delta_s = 0$. For a finite $\delta_s$, with the increase of $k$, $v^{DW}$ quickly increases to the maximal, estimated to be ~$1.414A/ds$ in the limit of $k \to \infty$. However, inside the domain, the group velocity, $v^{DO}$ gradually increases with increasing $k$. As a result, spin wave inside wall is much faster than that inside domain, especially in the low frequency regime.

Furthermore, we compute the energy consumptions for exciting spin waves inside the domain and wall respectively. The instantaneous power is written as $P(t) = \int_V \mathbf{M} \cdot \dot{\mathbf{h}}(t) d\mathbf{r}$ with $V$ the volume of microwave field sources, $\mathbf{M}$ the local magnetic moment in the volume, and $\dot{\mathbf{h}}(t)$ the derivative of the field with respect to *t*. Generally, the energy consumption to excite spin waves inside wall is much lower than inside domain. For $\delta_s = 0$ and $f = 150$ GHz, as an example, the excitation power reaches up to 224.35 nW inside domain, while only ~13.25 nW inside wall. Notably, the spin-wave excitation and propagation inside the wall not only has a larger velocity, but also consumes a much lower energy. Thus, spin-wave channeling inside the wall significantly facilitates the spin wave transmission.

Finally, it is noted that current-induced spin-wave Doppler shift inside domain for a compensated ferrimagnets may appear, stimulating a supplementary investigation here for the Doppler shift of spin waves inside a FiM wall. Based on Eq. (8), the spin-wave Doppler shift $\Delta\omega_\pm$ can be estimated by $\Delta\omega_\pm = \omega_\pm - \omega_{\pm,0}$ with the frequency under zero current $\omega_{\pm,0} = \left(\pm\delta_s + \sqrt{\delta_s^2 + 4\rho A k^2}\right)/2\rho$. Then, $\Delta\omega_\pm$ is given by

$$\Delta\omega_\pm = \frac{s b_j^- k}{\sqrt{\delta_s^2 + 4\rho A k^2}}, \tag{12}$$

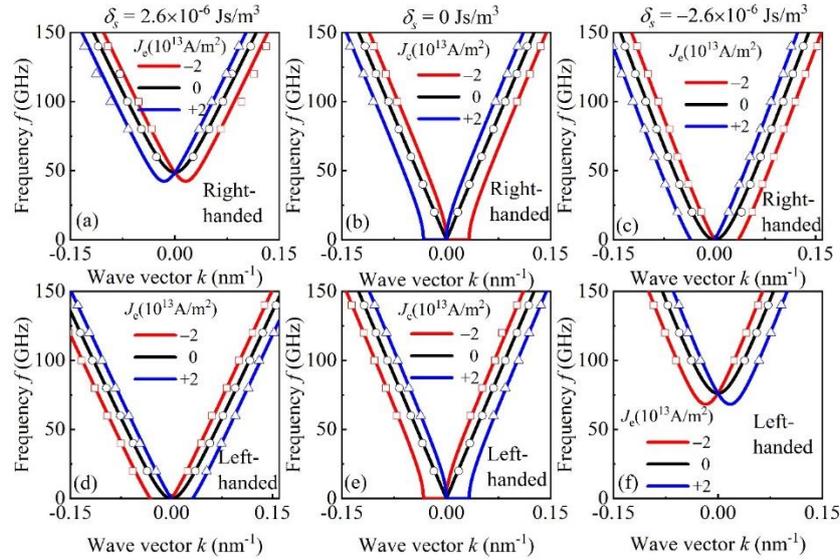

Fig. 4 (color online). Current-induced Doppler shift of right-handed spin waves in a FiM system for (a) $\delta_s > 0$, (b) $\delta_s = 0$, and (c) $\delta_s < 0$, and the shift of left-handed spin waves for (d) $\delta_s > 0$, (e) $\delta_s = 0$, and (f) $\delta_s < 0$.

Eq. (12) shows that the shift sign depends on the electric current, spin-wave handedness, and wave vector. For example, for the right-handed spin wave and $k > 0$, a positive (negative) current with $b_{J,i} < 0$ ($b_{J,i} > 0$) increases (decreases) the spin wave frequency. This behavior is confirmed in our simulations, and the corresponding results are given in Fig. 4 (a) where the calculated (lines) and simulated (symbols) dispersions for the right-handed spin waves as $\delta_s > 0$ are plotted. The simulated results coincide well with the theoretical calculations, demonstrating again the validity of the theory. Moreover, the sign of the Doppler shift is independent of $\delta_s$, as shown in Figs. 4(b) and 4(c) where the dispersions for $\delta_s = 0$ and $\delta_s < 0$, respectively, are plotted.

On the other hand, the sign of the Doppler shift of the left-handed spin wave is opposite to that of the right-handed spin wave, as demonstrated in Figs. 4(d)-4(f). Specifically, for $k > 0$, a positive (negative) current decreases (increases) the spin wave frequency. Thus, the Doppler shift of the spin waves in the FiM wall is similar to that in the uniform state.

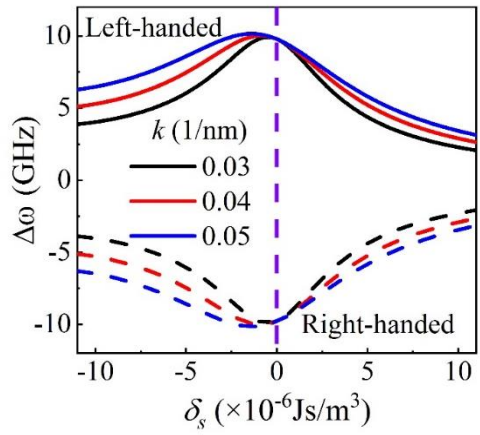

Fig. 5 (color online). Doppler shift for right- (dashed lines) and left-handed (solid lines) spin waves as a function of $\delta_s$ for various wave vectors.

A quantitative prediction of the Doppler shift is given in Fig. 5, where the calculated $\Delta\omega$ for the right- and left-handed spin waves as a function of $\delta_s$ for various $k$ is plotted, demonstrating the significant dependence of $\Delta\omega$ on $\delta_s$ value. In the vicinity of $T_A$ for $\delta_s \sim 0$, $\Delta\omega$ is estimated to be $\Delta\omega = sb_j^-/4\rho A$, independent of wave vector $k$. Furthermore, the maximum of the Doppler shift appears at $\delta_s < 0$, and the peak position shifts toward high $|\delta_s|$ side as $k$ increases. To some extent, the dependence of $\Delta\omega$ on $\delta_s$ could allow one to estimate the net angular momentum by measuring the current-induced spin-wave Doppler shift in future experiments.

In conclusion, we have studied theoretically and numerically the excitation and propagation of spin waves in ferrimagnetic domain wall. Our study shows that the spin-wave dispersion depends on both the net angular momentum $\delta_s$ and the wave handedness. For a positive $\delta_s$, the dispersion of the left-handed spin waves is gapless, while that of the right-handed spin waves is still with a forbidden frequency gap. Thus, it is suggested that FiM domain wall could serve as a filter in which only spin wave with specific handedness can be excited and propagated. Importantly, spin-wave dynamics inside wall is much faster than inside domain and the excitation consumes much less energy, demonstrating again the advantages of domain walls serving as waveguides. Moreover, the current-induced spin-wave Doppler shift is also revealed inside wall. It is demonstrated that $\delta_s$ can effectively modulate the frequency shift magnitude, like that inside domain. Thus, this work unveils interesting spin-wave dynamics in FiM walls, which is helpful in future experiment design and spin wave applications.

**Acknowledgment**

We sincerely appreciate the insightful discussions with Lingling Song. The work is supported by the Natural Science Foundation of China (Grant Nos. 51971096, 92163210, 51721001), and the Natural Science Foundation of Guangdong Province (Grant No. 2019A1515011028), and the Science and Technology Planning Project of Guangzhou in China (Grant No. 201904010019).